\documentstyle[12pt]{article}
\let\section=\subsection     \let\subsection=\subsubsection                
\newcommand{\be}{\begin{equation}}
\newcommand{\ee}{\end{equation}}
\newcommand{\bea}{\begin{eqnarray}}
\newcommand{\eea}{\end{eqnarray}}

\begin{document}
\begin{center}
   {\large \bf DILEPTONS AND HADRON SPECTRAL}\\[2mm]
   {\large \bf FUNCTIONS IN HEAVY-ION COLLISIONS\footnote{supported by GSI Darmstadt}}\\[5mm]
   W.~CASSING\footnote{In collaboration with E. L. Bratkovskaya and S. Juchem} \\[5mm]
   {\small \it  Institut f\"ur Theoretische Physik, Universit\"at Giessen\\
D-35392 Giessen, Germany \\[8mm] }
\end{center}

\begin{abstract}\noindent
We briefly review the dilepton experiments at BEVALAC/SIS and SPS
energies for $p+p$, $p+A$ and $A+A$ collisions as well as our
present understanding of the data within transport theoretical
simulations. Since dileptons from $p+A$ and $A+A$ collisions in
particular probe the in-medium spectral functions of vector
mesons, a novel semiclassical off-shell transport approach is
introduced on the basis of the Kadanoff-Baym equations that
describes the dynamical evolution of broad hadron spectral
functions. The implications of the off-shell dynamics -- relative
to the conventional on-shell transport dynamics -- are discussed
for proton spectra, high energy $\gamma$, pion, kaon and antikaon
production from GANIL to AGS energies in comparison to
experimental data.
\end{abstract}

\section{Introduction}
The properties of hadrons in hot and/or dense nuclear matter are
of central interest for the nuclear physics community as one
expects to learn about precurser effects for chiral symmetry
restoration or to explore the vicinity of a quark-gluon plasma
(QGP) phase transition. Whereas the early 'big bang' most likely
evolved through equilibrium configurations from the QGP to a
hadronic phase, this is not the case for the hot/dense systems
produced in collisions of heavy ions at relativistic energies.

Nowadays, the dynamical description of strongly interacting
systems out of equilibrium is dominantly based on transport
theories and efficient numerical recipies have been set up for the
solution of the coupled channel transport equations
\cite{URQMD,ca99} (and Refs. therein). These transport approaches
have been derived either from the Kadanoff-Baym equations
\cite{kb62} or from the hierarchy of connected equal-time Green
functions \cite{Wang1,Zuo} by applying a Wigner transformation and
restricting to first order in the derivatives of the phase-space
variables ($X, P$). Whereas theoretical formulations of off-shell
quantum transport have been limited to the formal level for a
couple of years \cite{Bot,ph95} only recently a tractable
semiclassical form has been derived for testparticles in the eight
dimensional phase-space of a particle \cite{ca1,ca2}.

In this contribution a brief review is given in Section 2 on the
present understanding of 'low mass' dilepton data from $pp$ to
$AA$ collisions and the necessity for a quantum transport
description is pointed out. A short reminder of the steps for a
derivation of such transport theories is presented in Section 3 as
well as generalized testparticle equations of motion that are
extracted in the semiclassical limit. Section 4 is devoted to a
presentation of the most important off-shell effects in
nucleus-nucleus collisions from GANIL to AGS energies.

\section{Low mass dileptons}
Since the first dilepton studies from nucleus-nucleus collisions
at the BEVALAC by the DLS Collaboration \cite{Roche} the field of
dilepton measurements has rapidly expanded at CERN/SPS (see Refs.
\cite{ca99,Rapp99}); the new detector HADES will complement the
experimental programm at the SIS \cite{Hades}. The data taken by
the DLS Collaboration \cite{DLS} on the elementary $pp$ reaction
can reasonably be well described by the production of intermediate
baryonic resonances as well as $\pi^0, \eta, \omega, \rho^0$ and
$\phi$ mesons and their Dalitz or direct decays to $e^+e^-$ pairs.
This has been demonstrated in detail in Refs. \cite{brat99,Ernst}.
A similar statement holds true for the $p + Be$ and $p + W$
reactions at 450 GeV and 200 GeV \cite{ca99}, respectively, which
are fully described by the meson Dalitz or direct decays to
dileptons since baryon resonance decays play no longer a
substantial role at SPS energies. Whereas the $e^+e^-$ and
$\mu^+\mu^-$ differential spectra of the CERES and HELIOS-3
Collaborations -- that show an excess of dileptons in the
invariant mass regime 0.3 GeV $\leq M \leq $ 0.6 GeV -- can be
described within the 'dropping mass' scenario \cite{Gerry} or the
'melting' $\rho$-meson scenario \cite{Rapp99,Rapp97} -- that
involves a strong broadening of the $\rho$ spectral function in
the medium due to the coupling to dressed pions and resonance-hole
loops -- the latter concepts seem to work no longer for the DLS
data at 1 A GeV as worked out by Bratkovskaya et al.
\cite{brat98,brat99a}. Here, the $e^+e^-$ invariant mass spectra
for $Ca + Ca$ are underestimated in the regime 0.2 GeV $\leq M
\leq$ 0.6 GeV by a factor of 6--7 when involving 'vacuum' spectral
functions for the $\rho$ and $\omega$ mesons \cite{brat98} and by
a factor of 2--3 within the 'dropping mass' \cite{brat98} and
'melting' $\rho$ scenarios \cite{Ernst,brat99a} (DLS-puzzle). In
part this discrepancy might be attributed to an improper 'use' of
the vector-dominance-model (VDM) in elementary reactions or
unknown isospin dependencies in reactions involving neutrons as
pointed out by Mosel at this workshop and should be examined
experimentally first in $\gamma p$ or $\pi^-p$ and $\pi^-d$
reactions \cite{Mosel}. These elementary reactions also are
expected to provide valuable insight into $\rho/\omega$ mixing and
their individual production processes \cite{Soyeur}.

A general survey on low mass dilepton production in $Au + Au$
collisions from SIS to RHIC energies has been given in Refs.
\cite{ca99,ca3}; here the energy regime from 2 -- 10 A GeV is
found to be most promising for studies of the vector meson
spectral functions at high baryon density since the systems show
high density regimes above $2-3$ $\rho_0$ for more than 10 fm/c
which are large compared to the $\rho$ and $\omega$ life times in
the medium. Furthermore, at a couple of A GeV the scalar quark
condensate $<\bar{q} q>$ is expected to approach zero, i.e. a
chirally restored phase, for substantial space-time volumes
\cite{ca3,Friman}. At SPS energies and above most vector mesons
are produced dominantly by meson-meson channels in the
longitudinally expanding fireball at rather low baryon density
(but high temperature) \cite{ca3}. This lowers the perspectives of
low mass dilepton measurements at RHIC energies; however,
intermediate mass dileptons from 1.2 -- 2.5 GeV of invariant mass
might show a signal from the QGP phase provided that open charm is
strongly suppressed when propagating through a colored QGP medium
\cite{Rapp2000}.

In view of these more general considerations an accurate
understanding of the vector meson properties in the medium -- or
the imaginary part of the current-current correlation functions
\cite{Weise} -- can only be achieved if the full dynamical
evolution of the hadron spectral functions -- also for
nonequilibrium phase-space configurations -- can be followed
throughout all stages of a nucleus-nucleus collision. The next
Section is devoted to a formulation of such type of off-shell
transport theory following Refs. \cite{ca1,ca2}.

\section{Extended semiclassical transport equations}
The general starting point for the derivation of a transport
equation for particles with a finite and dynamical width are the
Dyson-Schwinger equations for the retarded and advanced Green
functions $S^{ret}$, $S^{adv}$ and for the non-ordered Green
functions $S^{<}$ and $S^{>}$ \cite{ca1}. In the case of scalar
bosons -- which is considered in the following for simplicity --
these Green functions are defined by
\begin{equation}  i \, S^{<}_{xy}  :=  \; < \, \Phi^{\dagger}(y) \; \Phi(x)
\,
> \, , \hspace{1cm} i \, S^{>}_{xy}
 :=  \; <\,  \Phi(x) \; \Phi^{\dagger}(y) \, > \, , \end{equation}
 $$
 i  S^{ret}_{xy}  :=  \Theta (x_0 - y_0) \  <
 [  \Phi(x) , \Phi^{\dagger}(y) ]
>  , \  i \, S^{adv}_{xy}  :=  -  \Theta (y_0 -
x_0) \, <  [ \Phi(x) , \Phi^{\dagger}(y) ] \, >  . $$
\noindent
 They depend on the space-time coordinates $x,y$ as
indicated by the indices $\cdot_{xy}$. The Green functions are
determined via Dyson-Schwinger equations by the retarded and
advanced self energies $\Sigma^{ret},\Sigma^{adv}$ and the
collisional self energy $\Sigma^{<}$:
 \bea \hat{S}_{0x}^{-1} \; S_{xy}^{ret}
=  \: \delta_{xy} \; +  \: \Sigma_{xz}^{ret} \: \odot \:
S_{zy}^{ret} \; , \ \
 \hat{S}_{0x}^{-1} \; S_{xy}^{adv}  \: =  \: \delta_{xy}
 \: +  \: \Sigma_{xz}^{adv} \: \odot \: S_{zy}^{adv}  ,
\label{dsadv_spatial} \eea
\bea \hat{S}_{0x}^{-1} \; S_{xy}^{<}  \: =  \: \Sigma_{xz}^{ret}
\: \odot \: S_{zy}^{<}  \: +  \: \Sigma_{xz}^{<} \: \odot \:
S_{zy}^{adv} \: , \label{kb_spatial} \eea
where Eq. (\ref{kb_spatial}) is the well-known Kadanoff-Baym
equation. Here $\hat{S}^{-1}_{0x}$ denotes the (negative)
Klein-Gordon differential operator which is given for bosonic
field quanta of (bare) mass $M_0$ by $\hat{S}^{-1}_{0x} = -
(\partial^{\mu}_{x} \partial^{x}_{\mu} + M^2_0 )$; $\delta_{xy}$
represents the four-dimensional $\delta$-distribution $\delta_{xy}
\equiv \delta^{(4)}(x-y)$ and the symbol $\odot$ indicates an
integration (from $-\infty$ to $\infty$) over all common
intermediate variables (cf. \cite{ca1}).

For the derivation of a semiclassical transport equation one now
changes from a pure space-time formulation into the
Wigner-representation with
variable $X = (x+y)/2$ and the four-momentum $P$,
which is introduced by Fourier-transformation with respect to the
relative space-time coordinate $(x-y)$. In any semiclassical
transport theory one, furthermore,
keeps only contributions up to the first order in the space-time gradients.
To simplify notation the operator $\Diamond$ is introduced as \cite{ph95,ca1}
\bea
 \Diamond \, \{ \, F_{1} \, \} \, \{ \, F_{2} \, \}
\; := \; \frac{1}{2} \left( \frac{\partial F_{1}}{\partial
X^{\mu}} \: \frac{\partial F_{2}}{\partial P_{\mu}} \; - \;
\frac{\partial F_{1}}{\partial P_{\mu}} \: \frac{\partial
F_{2}}{\partial X^{\mu}} \right) ,\label{poissonoperator} \eea
which is a four-dimensional generalization of the well-known
Poisson-bracket. In first order gradient expansion one then
obtains algebraic relations between the real and the imaginary
part of the retarded Green functions. On the other hand Eq. (3)
leads to a 'transport equation' for the Green function $S^{<}$. In
order to obtain real  quantities one separates all retarded and
advanced functions -- Green functions and self energies  -- into
real and imaginary parts,
\bea S_{XP}^{ret,adv} \; = \;
Re S^{ret}_{XP} \; \mp \; \frac{i}{2} \, A_{XP} \; , \qquad
\Sigma_{XP}^{ret,adv} \; = \; Re \Sigma^{ret}_{XP} \; \mp \;
\frac{i}{2} \, \Gamma_{XP} \; . \label{ret_sep} \eea
The imaginary part of the retarded propagator is given (up to a
factor 2) by the normalized spectral function
\bea A_{XP} \: = \: i \left[ \, S_{XP}^{ret} \: - \: S_{XP}^{adv}
\, \right] \: = \: - 2 \, Im \, S^{ret}_{XP} \; ,  \qquad \int
\frac{d P_0^2}{4 \pi} \,  A_{XP} \; = \; 1 \; ,
\label{spectralfunction} \eea
while the imaginary part of the self energy corresponds to half
the width $\Gamma_{XP}$. By separating the complex equations into
their real and imaginary contributions one obtains an algebraic
equation between the real and the imaginary part of $S^{ret}$ (Eq.
(12) of \cite{ca1}),  an algebraic solution for the spectral
function (in first order gradient expansion) as
\bea A_{XP} \; = \; \frac{ \Gamma_{XP} } {( \, P^2 \, - \,
M_{0}^{2} \, - \, Re \Sigma^{ret}_{XP} )^{2} \: + \:
\Gamma_{XP}^{2}/4} \; , \label{alg_spectral} \eea
as well as for the real part of the retarded propagator
\cite{ca1}.

The (Wigner-transformed) Kadanoff-Baym equation (3) allows for the
construction of a transport equation for the Green function
$S^{<}$. When separating the real and the imaginary contribution
of this equation one obtains i) a generalized transport equation
(Eq. (15) in \cite{ca1}) and ii) a generalized mass-shell
constraint (Eq. (16) in \cite{ca1}). Furthermore, according to
Botermans and Malfliet \cite{Bot}, in the transport equation the
collisional self energy $\Sigma^{<}$ has to be replaced by $S^{<}
\cdot \Gamma / A$ to gain a consistent first order gradient
expansion scheme. Finally, the general transport equation (in
first order gradient expansion) reads \cite{ca1}
$$ A_{XP} \, \Gamma_{XP} \  [ \, \Diamond \; \{ \, P^2 - M_0^2 -
Re \Sigma^{ret}_{XP} \, \} \; \{ \, S^<_{XP} \, \} $$
\begin{equation} - \: \frac{1}{\Gamma_{XP}} \; \Diamond \; \{ \, \Gamma_{XP}
 \} \; \{ \, ( \, P^2 - M_0^2 - Re \Sigma^{ret}_{XP} \, ) \,
S^<_{XP} \, \} \, ]  = \; i \, \left[ \, \Sigma^>_{XP} \: S^<_{XP}
\: - \: \Sigma^<_{XP} \: S^>_{XP} \, \right]. \label{trans_approx}
\end{equation}
It has also been independently derived by Ivanov et al. \cite{Knoll} and
Leupold \cite{Leupold}.
Its formal structure is fixed by the approximations applied. Note,
however, that the dynamics is fully determined by the different
self energies, i.e. $Re \Sigma_{XP}^{ret}, \Gamma_{XP},
\Sigma_{XP}^<$ and $\Sigma_{XP}^>$ that have to be specified for
the physical systems of interest.

Besides the drift term (i.e. $\Diamond \{P^2 - M^2_0\} \{ S^{<} \}
= - P^{\mu} \partial^{X}_{\mu} S^{<} )$ and the Vlasov term (i.e.
$-\Diamond \{ Re \Sigma^{ret} \} \{ S^{<} \} $) a third
contribution appears on the l.h.s. of (8) (i.e. $\sim \Diamond
\{\Gamma_{XP}\} \{ \cdots S_{XP}^{<} \} $), which vanishes in the
quasiparticle limit and incorporates -- as shown in
\cite{ca1,ca2,Leupold} -- the off-shell behaviour in the particle
propagation that has been neglected so far in transport
studies\footnote{This also holds true for the recent numerical
studies in Ref. \cite{Effe}}. The r.h.s. of (8) consists of a
collision term with its characteristic gain ($\sim \Sigma^{<}
S^{>}$) and loss ($\sim \Sigma^{>} S^{<}$) structure, where
scattering processes of particles into and out of a given
phase-space cell are described.

\subsection{Testparticle approximation}
In order to obtain an approximate solution to the transport
equation (\ref{trans_approx})  a testparticle ansatz is used for
the Green function $S^{<}$, more specifically for the real and
positive semidefinite quantity $F_{XP} \; = A_{XP} N_{XP} = \; i
\, S^{<}_{XP} \;$,
\bea F_{XP} \; \sim \; \sum_{i=1}^{N} \; \delta^{(3)} ({\vec{X}}
\, - \, {\vec{X}}_i (t)) \; \; \delta^{(3)} ({\vec{P}} \, - \,
{\vec{P}}_i (t)) \; \; \delta(P_0 - \, \epsilon_i(t)) \: .
\label{testparticle} \eea
In the most general case (where the self energies depend
on four-momentum $P$, time $t$ and the spatial coordinates
$\vec{X}$) the equations of motion for the testparticles  read \cite{ca2}
\begin{equation}
 \label{eomr} \frac{d {\vec X}_i}{dt} \! = \  \frac{1}{1 - C_{(i)}} \, \frac{1}{2 \epsilon_i} \: \left[ \, 2
\, {\vec P}_i \, + \, {\vec \nabla}_{P_i} \, Re \Sigma^{ret}_{(i)}
\, + \, \frac{ \epsilon_i^2 - {\vec P}_i^2 - M_0^2 - Re
\Sigma^{ret}_{(i)}}{\Gamma_{(i)}} \: {\vec \nabla}_{P_i} \,
\Gamma_{(i)} \: \right],
\end{equation}
\begin{equation}
\label{eomp} \frac{d {\vec P}_i}{d t} \!  =  \! -
\frac{1}{1-C_{(i)}} \, \frac{1}{2 \epsilon_{i}} \: \left[ {\vec
\nabla}_{X_i} \, Re \Sigma^{ret}_i \: + \: \frac{\epsilon_i^2 -
{\vec P}_i^2 - M_0^{2} - Re \Sigma^{ret}_{(i)}}{\Gamma_{(i)}} \:
{\vec \nabla}_{X_i} \, \Gamma_{(i)} \: \right], \end{equation}
\begin{equation}
\label{eome} \frac{d \epsilon_i}{d t} \!
 =  \!  \frac{1}{1 -
C_{(i)}} \, \frac{1}{2 \epsilon_i} \: \left[ \frac{\partial Re
\Sigma^{ret}_{(i)}}{\partial t} \: + \: \frac{\epsilon_i^2 - {\vec
P}_i^2 - M_0^{2} - Re \Sigma^{ret}_{(i)}}{\Gamma_{(i)}} \:
\frac{\partial \Gamma_{(i)}}{\partial t} \right],
\end{equation}
where the notation $F_{(i)}$ implies that the function is taken at
the coordinates of the testparticle, i.e. $F_{(i)} \equiv
F(t,\vec{X}_{i}(t),\vec{P}_{i}(t),\epsilon_{i}(t))$. These
equations also have been derived by Leupold  in the
nonrelativistic limit recently \cite{Leupold}.

In (\ref{eomr}-\ref{eome}) a common multiplication factor
$(1-C_{(i)})^{-1}$ appears, which contains the energy derivatives
of the retarded self energy
\bea \label{correc} C_{(i)} \: = \: \frac{1}{2 \epsilon_i} \left[
\frac{\partial}{\partial \epsilon_i} \, Re \Sigma^{ret}_{(i)} \: +
\: \frac{\epsilon_i^2 - {\vec P}_i^2 - M_0^2 - Re
\Sigma^{ret}_{(i)}}{\Gamma_{(i)}} \: \frac{\partial }{\partial
\epsilon_i} \, \Gamma_{(i)} \right] \:  \eea
and yields a shift of the system time $t$ to the 'eigentime' of
particle $i$ defined by $\tilde{t}_{i} = t /(1-C_{(i)})$
\cite{ca2}. As the reader immediately verifies, the derivatives
with respect to the 'eigentime', i.e. $d \vec{X}_i / d
\tilde{t}_i$, $d \vec{P}_i / d \tilde{t}_i$ and $d \epsilon_i / d
\tilde{t}_i$ then emerge without this renormalization factor for
each testparticle $i$ when neglecting higher order time
derivatives in line with the semiclassical approximation scheme.

For momentum-independent self energies one regains the transport
equations as derived in \cite{ca1}. Furthermore, in the limiting
case of particles with vanishing gradients of the width
$\Gamma_{XP}$ these equations of motion  reduce to the well-known
transport equations of the quasiparticle picture.

Furthermore, the variable $M^{2} = P^2 - Re \Sigma^{ret}$ is taken
as an independent variable instead of $P_0$. Eq. (\ref{eome}) then
turns to
\bea \label{eomm} \frac{dM_i^2}{dt} \; = \; \frac{M_i^2 -
M_0^2}{\Gamma_{(i)}} \; \frac{d \Gamma_{(i)}}{dt} \eea
for the time evolution of the testparticle $i$ in the invariant
mass squared \cite{ca1,ca2}.

\subsection{Collision terms}
The collision term of the Kadanoff-Baym equation can only be
worked out by giving explicit approximations for $\Sigma^{<}$ and
$\Sigma^{>}$. As in the conventional transport theory the
formulation of collision terms is based on Dirac-Brueckner theory
for the transition amplitudes. The formal structure resembles
somewhat the on-shell formulations \cite{ca99}, however, includes
additionally the spectral functions of all hadrons in the initial
and final channels and four-momentum integrations instead of
three-momentum integrals \cite{ca2}. Furthermore, now all
off-shell transition amplitudes enter that so far are scarcely
known for strong interaction processes. The collisional width
$\Gamma_{coll}(X,{\vec P},M^2)$ for each hadron then is defined
via the loss-term of the corresponding collision integral
\cite{ca2}.

For the numerical studies mentioned below the off-shell transition
amplitudes squared have been taken as the on-shell values employed
in the conventional HSD approach \cite{ca99}, however, corrected
by the final phase-space for the particles with off-shell masses.
For a detailed description of the numerical recipies and
approximations employed the reader is refered to Refs.
\cite{ca1,ca2}.

\section{Off-shell effects in nucleus-nucleus collisions}
The off-shell propagation of hadrons in nucleus-nucleus collisions
has been examined from GANIL to AGS energies in Refs.
\cite{ca1,ca2} and for infinite nuclear matter problems in Ref.
\cite{ca4}. The main results are briefly summarized:
\begin{itemize}
\item{The numerical implementations of off-shell dynamics can be
shown to lead to the proper equilibrium mass distributions (for $t
\rightarrow \infty$) in line with the quantum statistical limit
when employing the same equilibrium spectral functions for all
hadrons \cite{ca4}.}
\item{Numerical studies of systems in a
finite box with periodic boundary conditions show that the
off-shell dynamics lead to practically the same equilibration
times as the on-shell dynamics \cite{ca4}.}
\item{At GANIL energies ($\sim$ 95 A GeV) the off-shell effects
(OSE) are most pronounced since the excitation of
$\Delta$-resonances is strongly suppressed. OSE show up especially
in the high momentum tails of the collisional $\sqrt{s}$
distribution, in high momentum proton spectra and most pronounced
in high energy photon production as demonstrated in comparison to
the $\gamma$ spectra from the TAPS collaboration for $Ar + Au$ at
95 A GeV \cite{ca1}.}
\item{At SIS energies (1--2 A GeV) the OSE lead only to a slight
enhancement of the high transverse momentum spectra since the high
mass tail of the nucleon spectral function remains small compared
to the $\Delta$ mass distribution \cite{ca2}. The latter again is
only marginally enhanced at high masses relative to the on-shell
dynamics which leads to almost the same pion spectra in this
energy domain. It's worth noting that pions in the high density
phase become somewhat harder, which leads to an enhanced
production of kaons and antikaons. The enhancement of $K^-$ for
$Ni + Ni$ reactions at 1.8 A GeV amounts to a factor of about 2
such that in-medium antikaon potentials should be substantially
less attractive than suggested in \cite{ca99}.}
\item{The 'DLS puzzle' mentioned in Section 2 might be solved
within the off-shell dynamics once there is a strong coupling of
low-mass $\rho^0$ mesons to baryons at high baryon density, which
approximately 'thermalizes' the $\rho$ mass distribution. Since
this is presently a {\it speculation}, no definite conclusions can
be drawn so far.}
\item{At AGS energies of 11 A GeV the particle production is
dominated by 'string' continuum excitations, which themselves are
similar to very broad hadron spectral functions. Since all
productions thresholds for $K,\bar{K}, \rho, \omega, \phi$ are by
far exceeded in the initial collisions, no substantial effects
from the off-shell dynamics could be established within the
numerical accuracy on rapidity distributions and transverse mass
spectra.}
\end{itemize}

\noindent In closing it is necessary to point out that although
the general equations for off-shell transport are available by now
and efficient numerical solution schemes exist, the final
understanding of this approach requires the knowledge of {\it all
off-shell (in-medium) transition amplitudes}! The present recipe
of using final state phase-space corrections is only a first step
on this way.

\end{document}